\documentclass[amsmath,amssymb,aps,prd,showpacs,nofootinbib]{revtex4}
\usepackage{graphicx}
\usepackage{amsfonts}
\usepackage{amssymb}
\usepackage{amsmath}
\usepackage{bm}
\usepackage{latexsym}

\newcommand{\beq}{\begin{eqnarray}}
\newcommand{\eeq}{\end{eqnarray}}
\newcommand{\be}{\begin{equation}}
\newcommand{\ee}{\end{equation}}

\def\fun#1#2{\lower3.6pt\vbox{\baselineskip0pt\lineskip.9pt
\ialign{$\mathsurround=0pt#1\hfil ##\hfil$\crcr#2\crcr\sim\crcr}}}

\newcommand{{\SD}}{\rm SD}

\newcommand{\vep}{\bm p}

\begin{document}

\title{Radial and orbital Regge trajectories in heavy quarkonia}

\author{\firstname{A.M.}~\surname{Badalian}}
\email{badalian@itep.ru} \affiliation{Institute of Theoretical and Experimental
Physics, Moscow, Russia}

\author{\firstname{B.L.G.}~\surname{Bakker}}
\email{b.l.g.bakker@vu.nl} \affiliation{Department of Physics and Astronomy,
Vrije Universiteit, Amsterdam, The Netherlands}

\date{\today}

\begin{abstract}
The spectra of heavy quarkonia  are studied in two approaches: with the use of the Afonin-Pusenkov representation of
the Regge trajectory for the squared excitation energy $E^2(nl)$ (ERT),  and using the relativistic Hamiltonian with the universal interaction.
The parameters of the ERTs are extracted from experimental mass differences and their values in bottomonium: the intercept $a(b\bar b)=0.131\,$GeV$^2$,
the slope of the orbital ERT $b_l(b\bar b) =0.50$\,GeV$^2$, and the slope of the radial ERT, $b_n(b\bar b)=0.724$\,GeV$^2$, appear to be smaller than those
in charmonium, where $a(c\bar c)=0.381$\,GeV$^2$, $b_l(c\bar c)=0.686$\,GeV$^2$, and the radial slope $b_n(c\bar c)= 1.074$\,GeV$^2$, which value is close
to that in light mesons, $b_n(q\bar q)=1.1(1)$\,GeV$^2$. For the resonances above the $D\bar D$ threshold the masses of the  $\chi_{c0}(nP)$ with
$n=2,3,4$, equal to 4218\,MeV, 4503\,MeV, 4754\,MeV, are  obtained, while above the $B\bar B$ threshold the resonances $\Upsilon(3\,^3D_1)$ with
the mass 10693\,MeV and $\chi_{b1}(4\,^3P_1)$ with the mass 10756\,MeV are predicted.
\end{abstract}

\maketitle

\section{Introduction}
In recent years  in heavy quarkonia (HQ) a large number of new resonances were observed
 \cite{1,2,3,4,5,6,7,8} and among them the resonances  $X(4500)$ and $X(4700)$ with $J^{PC}=0^{++}$ \cite{6} are
particularly interesting, being the highest excitations in the meson sector. The discovery of these resonances has stimulated new  theoretical
studies \cite{8,9,10} and different conceptions about their nature were presented, including
diquark-antidiquark $cs\,\bar c\bar s$ types of tetraquarks \cite{10,11,12,13}. However, even within the
tetraquark $cs\,\bar c\bar s$  picture  different interpretations were suggested.
Also the conventional $c\bar c$ structure of these resonances was studied \cite{14,15,16,17,18},
which implies that the $c\bar c$ component dominates in the wave function (w.f.) of a resonance,
but does not exclude that other components, like diquark-antidiquark or meson-meson, can also be present in the wave
function (w.f.) \cite{14}. For decades the spectra and other properties of HQ were studied in different potential models
(PMs), both non-relativistic and relativistic  \cite{19,20,21,22,23,24,25}, which allow  successfully to describe low-lying HQ states.
However, the masses of the high excitations  strongly depend on the $Q\bar Q$ interaction at large distances, as well as on the heavy quark mass used,
and their values can differ by $\sim (100-150)$~MeV (see the compilation in Ref.~\cite{23}). This happens because using in PMs
several fitting parameters, the first two or three excitations can be easily described with  a good accuracy,
while the masses of the high excitations appear to be very sensitive to behavior of the $Q\bar Q$ potential at large distances.
For example, in Ref.~\cite{18}, where the  screened confining potential is used, the resonance $X(4140)$ is
considered as a candidate to  $\chi_{c1}(3P)$,  while within a similar model the mass $M_{c1}(3P)$,
larger  by $\sim 140$\,MeV, is obtained \cite{14}, and this state is identified as $X(4274)$.

The spectra of HQ were also studied via  the radial Regge trajectories (RTs) with the parameters determined either in dynamical calculations \cite{21,23,25,26,27,28}, or in the analysis of the experimental masses \cite{29,30}, where high charmonium excitations are  described by  a linear radial  RT, similar to those found in light mesons \cite{31},
\be
 M^2(nl)= M^2(n=0,l) + \mu_{c\bar c}^2  n, ~~(n=n_r).
 \label{eq.1}
\ee
Here  the slope $\mu_{c\bar c}^2$ has a large value, $\mu^2(c\bar c)\sim (2.8-3.4)$\,GeV$^2$ \cite{19,21,25,26,27}
and  slightly  depends on the angular momentum $l$ \cite{21}. In bottomonium a larger slope $\mu^2(b\bar b)$
was obtained  \cite{21,27,28}, which lies in the range $(4-7)$\,GeV$^2$ for different sets of parameters of the
potential $V_0(r)$ (see Eq.~{\ref{eq.7}) and the $b$-quark mass $m_b$.

A different representation of high HQ excitations was suggested by  Afonin and  Pusenkov \cite{32,33}, who introduced a new
type of  radial RT in HQ, henceforth denoted as ERT, referring  to the squared  excitation energy $E^2(nl)$, which is defined as
$E(nl)=M(nl) -2 m_Q$.  Moreover, the authors assumed that there exists a universal radial ERT,
\be
 E^2(nl) = a  + b_n\, n,
\label{eq.2}
\ee
which can be applied to all unflavoured vector mesons, including $\rho$, $\phi$, charmonium, and bottomonium and
therefore  the mass $M(n\,^3S_1;q\bar q)$, given by
\be
 M(n\,^3S_1) = 2 m_Q  + \sqrt{a + b_n\, n},
\label{eq.3}
\ee
would have the same slope for these mesons and the values  $b_n=1.1$\,GeV$^2$ and $a=0.57$\,GeV$^2$ were  chosen in
Ref.~\cite{33}. As seen from  Eq.~(\ref{eq.3}), the HQ mass $M(n\,^3S_1)$ depends on the quark mass $m_Q$ taken; the values $m_c=1.17$\,GeV and
$m_b=4.33$\,GeV were taken there.

It also follows from  Eq.~(\ref{eq.3})  that  the ERT with the universal
slope $b$ and the intercept $a$ have equal mass differences,
\be
M(2\,^3S_1) - M(1\,^3S_1) = \sqrt{a  + b_n} - \sqrt{a};~~M(3\,^3S_1) - M(2\,^3S_1) = \sqrt{a + 2\,b_n} -\sqrt{a + b_n}, ~~\label{eq.4}
\ee
both  for the heavy and te light  vector mesons and a given radial quantum number $n=n_r$.
However, this statement  does not agree with the experimental values of the mass differences, which can differ by $\sim 100$~MeV (see Table~\ref{tab.1}). In
Table~\ref{tab.1} we give also the experimental numbers for the mass difference, $M(2\,^3P_1)-M(1\,^3P_1)$, which will be used later in the analysis of
the orbital ERT.
\begin{table}
\caption{The experimental mass differences (in MeV) in light mesons( $n\bar n$), charmonium, and
bottomonium\label{tab.1}}
\begin{tabular}{|c|c|c|c|}
\hline
   $\Delta$                          &   $n\bar n$      &  $c\bar c$    &   $b\bar b$ \\
\hline
$M(2\,^3S_1) - M(1\,^3S_1)$           &  690(25)         &   589(2)     &   563(1)\\

$M_{\rm cog}(2S)-M_{\rm cog}(1S)$    &   700(20)        &  605(2)    &   567(1) \\

$M(3\,^3S_1) - M(2\,^3S_1)$         &   415(55)         &  353(1)      & 332(1)   \\

$M(2\,^3P_1) - M(1\,^3P_1)$         &    424(49)         &  361(2)      & 363(1)  \\
\hline
\end{tabular}
\end{table}
From Table~\ref{tab.1} one can see that the mass difference between the first excited state and the ground state does not change, if instead of  the masses
$M(n\,^3S_1)$ one takes
the centroid masses $M_{\rm cog}(nS)$,  i.e., it does not depend on spin effects, being in light mesons larger than in charmonium and bottomonium,
which values that
differ only by 26\,MeV. Such close values of the mass differences could be partly explained by the existence of the universal potential,
 $V_0(r)=V_{\rm c}(r)+V_{\rm ge}$, which allows to describe the low-lying states of all mesons \cite{20,21,23,25}, however, this choice is not sufficient to
 obtain equal slopes of ERTs for heavy and light mesons (see below).

Notice that in HQ the mass formula is more simple than in a light meson, where it includes the self-energy and the string corrections
\cite{34,35}, which are small and can be neglected in HQ \cite{35}. However, the masses of heavy mesons, which have small
sizes,  are strongly affected by the gluon-exchange (GE) interaction and for them the asymptotic freedom (AF) behavior of the strong
coupling has to be taken into account, in contrast to light mesons, where the GE potential can be presented as
the Coulomb potential with the coupling $\alpha(\rm eff.)=const.$ and the AF behavior is important
only for the $1S$ and $1P$  states \cite{34}.

In the present  paper we study orbital and radial ERTs of HQ in the $(E^2,n)$ and $(E^2,nl)$ planes, having in mind three goals: (i) to extract the slope
of the radial RTs $\beta_n(Q\bar Q)$ from experiment; (ii) to determine the slope of orbital RTs
$\beta_l(Q\bar Q)$ for the HQ mesons and show that the slopes $\beta_n$ and $\beta_l$ are different
in charmonium  and bottomonium and smaller than those in light mesons; (iii) to show that the generalized  ERT with the $E^2(nl) = a + b_l \,l + b_n\, n$, with  different slopes $b_l, b_n$  can be introduced  in charmonium and bottomonium.

\section{The Regge trajectories in the ($E^2,n$) and ($E^2,l$) planes in bottomonium}

Bottomonium has a large number of levels below the open flavour threshold and provides the unique possibility
to extract the parameters of the ERT in the ($E^2,n$)- and ($E^2,l$)-planes from experiment with high accuracy.
For that it is sufficient to use the mass differences (see Eqs.~\ref{eq.4}),
$M(\Upsilon(2S))- M(\Upsilon(1S))=\sqrt{a + b_n}-\sqrt{a}=0.563(1)$~MeV and $M(\Upsilon(3S)) -
M(\Upsilon(2S)=\sqrt{a+2 b_n} -\sqrt{a+b_n}=0.355(5)$~MeV and also the definition of the ground state mass,
$M(\Upsilon(1S)=\sqrt{a} + 2  m_b = 9.460(1)$~GeV.
From these relations the following values of the slope and the intercept of the radial ERT in
the ($E^2,n$)-plane  are calculated,
\be
 a = 0.1307\,{\rm GeV}^2,\quad b_n (l=0)= 0.7242\,{\rm GeV}^2.
 \label{eq.5}
\ee
Note, that the slope $b_n(\Upsilon)$ appears to be smaller than the slope of the $\rho(n\,^3S_1)$ trajectory,
$b_n(\rho)=\mu^2\approx 1.45$\,GeV$^2$, which follows from the experimental values of the masses $M(\rho(nS))$  \cite{1}. As the next
important step, knowing the intercept $a$ and the ground state mass $M(\Upsilon(1S))=9.460$\,GeV, we extract the quark mass $m_b$,
\be
 m_b = 4.5492\,{\rm GeV}.
 \label{eq.6}
\ee
This mass appears to be  not a fitting parameter, but just coincides with the one-loop pole mass
$m_b(1-{\rm  loop})=1.086\, \bar{m}_b=4.550$\,GeV, if the conventional current mass $\bar{m}_b=4.18(1)$\,GeV
and the QCD  constant $\Lambda_{\overline{MS}}(n_f=5)=200$\,MeV (or $\alpha_s(\bar{m}_b)=0.20$)
\cite{36,37,38} are adopted.

The masses of $\Upsilon(nS)$, defined by the radial ERT, Eq~(\ref{eq.3}) with the parameters from
Eqs.~(\ref{eq.5},\ref{eq.6}), are  presented in Table~\ref{tab.2}, where also the masses of the same states, calculated
with the spinless Salpeter equation (SSE),
\be
 \left( 2 \sqrt{\vep^2 + \tilde{m}^2_b} + V_0(r) \right) \varphi_{nl}(r) = M_{\rm cog}(nl) \varphi_{nl}(r),
\label{eq.7}
\ee
are given. Notice that in this SSE the two-loop mass  $\tilde{m}_b=4.830(5)$\,GeV enters and the
static potential $V_0(r)= \sigma r  - \frac{4\alpha_{\rm V}(r)}{3 r}$ is defined by  the parameters from Ref.~\cite{37},
\be
 \tilde{m}_b=4.830\,{\rm GeV}, ~ \sigma=0.18\,{\rm GeV}^2, ~ \Lambda_{\rm V}(n_f=3)=0.480\,{\rm GeV},
 ~M_{\rm B}=2\pi \sigma\approx 1.15\,{\rm GeV},
\label{eq.8}
\ee
which are not fitting parameters but defined on fundamental grounds and therefore the eigenvalues (e.v.s) $M_{\rm cog}(nl)$ of Eq.~(\ref{eq.7}) do not depend on any fitting parameters.  Only one extra parameter, $\alpha_{\rm hf}$, is present in the  hyperfine correction to the masses
$M(\Upsilon(nS)) = M_{\rm cog}(nS) + 1/4\,\delta_{\rm hf}$ (for $\delta_{\rm hf}$ see Ref.~\cite{39}).
\begin{table}
\caption{The experimental masses $M(\Upsilon(nS))$ (in MeV) \cite{1}, the masses $M(\Upsilon(nS))$, defined by the ERT, Eq.~(\ref{eq.2}), with the parameters from Eqs.~(\ref{eq.5},\ref{eq.6}), and the solutions of  Eq.~(\ref{eq.7}), $M(\Upsilon(nS)= M_{\rm cog}(nS) +1/4\,\delta_{\rm hf}(nS)$}\label{tab.2}
\begin{tabular}{|c|c|c|c|}
\hline
State   &      from ERT   & $M_{\rm cog}(nS) + 1/4\,\delta_{\rm hf}$ & experiment \cite{1}  \\
\hline
$\Upsilon(1S$     &  9460   &  9465      &   9460.3(3)    \\

$\Upsilon(2S)$     & 10023   &  10017   &    10023.3(3)\\

$\Upsilon(3S)$    & 10355  &    10359     &  10355.2(5) \\

$\Upsilon(4S)$   &  10616    &  10635     &    10579.4(1.2)\\

$\Upsilon(5S)$    &  10838  & 10884       &    10891 (4)\\

$\Upsilon(6S)$    &  11035    & 11093     &   $10987^{+11}_{-3} $ \\

$3\,^3D_1$    &  10693    &  10701   &  abs.   \\

$4\,^3D_1$    &   10901   &  10933    & abs.  \\
\hline
\end{tabular}
\end{table}

From Table~\ref{tab.2} one can see that the masses of $\Upsilon(nS)$ with $n=0,1,2$ are exactly
equal to the experimental values,  while the masses of the states with $n=3,4,5$,
which lie above the $B\bar B$ threshold, have mass shifts, e.g. the  $\Upsilon(4S)$ is shifted down by
38~MeV. The  situation with $\Upsilon(5S)$ and $\Upsilon(6S)$ is more complicated, because the ERT
gives the  mass of $\Upsilon(5S)$ by $\sim 40$~MeV smaller than the experimental value,
i.e.,  this resonance does not show the typical  mass shift up. This may occur
because of the nearby located  $3\,^3D_1$ and $4\,^3D_1$ resonances (see Table~\ref{tab.2} and the
calculations below), and also because the threshold $B_s\bar B_s$ (its mass $M_{\rm thres.}=10831(1)$ \,MeV) is close by.
Thus, here we face the channel-coupling problem, where a shift of one resonance up and of another  resonance down is possible.
The same many-channel situation takes place in the region near 11 GeV, where a mass shift down of the
$\Upsilon(6S)$ is possible due to the $S-D$ mixing of $\Upsilon(6S)$ and $\Upsilon(4D)$ \cite{40}.

To describe the orbital excitations we introduce the generalized ERT,
\be
 E^2(nl) = a + b_n\, n + b_l\, l,
\label{eq.9}
\ee
where the parameters $a$ and $b_n$, as well as $m_b$, are already defined and given in
Eqs.~(\ref{eq.5},\ref{eq.6}). To find the slope $b_l$ one can use the experimental mass of $\chi_{b1}(1P)$ (with $l=1,n-0$):
$M(\chi_{b1}(1P)) =9.893(1)$\,GeV$ = 2 m_Q + \sqrt{a+b_l}$. It gives $b_l(b\bar b)=0.50$\,GeV$^2$.

When higher excitations with $l\not= 0$ are considered in bottomonium,
the radial slope, extracted from the mass difference $M(\chi_{b1}(2P))- M(\chi_{b1}(1P))=0.362$~GeV, appears to be a bit smaller than that in the $\Upsilon(nS)$-family, namely,
$\beta_n(l\not= 0)=0.7060$\,GeV$^2$. This situation is similar to the one in light mesons, where the radial slope of the $\rho(nS)$-trajectory is larger than that for
$a_1(nP)$ and the $\rho(nD)$ mesons.

Then the complete set of  parameters of the generalized ERT, Eq.~(\ref{eq.9}) is:
\be
 a(b\bar b)=0.1307\,{\rm GeV}^2, ~~b_n(b\bar b,l=0)= 0.7242\,{\rm GeV}^2,~~b_n(l\not= 0)=0.7060\,{\rm GeV}^2, ~b_l(b\bar b) = 0.50\,{\rm GeV}^2, ~m_b=4.5492\,{\rm GeV}.
 \label{eq.9a}
\ee
The masses of  $\chi_{b1}(nP)$ and $\Upsilon(n\,^3D_1)$, calculated with the use of this  ERT,
Eq.~(\ref{eq.9},\ref{eq.9a}), are given in Table~\ref{tab.3}.
\begin{table}
\caption{The masses of the $\chi_{b1}(nP)$ and $\Upsilon(n\,^3D_1)$ (in MeV), calculated from ERT Eq.~\ref{eq.8} and the solutions of the SSE \ref{eq.4}}
\label{tab.3}

\begin{tabular}{|c|c|c|c|}
\hline
  State            &       ERT     &   Solutions of SSE    &   experiment \cite{1}\\
\hline
  $\chi_{b1}(1P)$  &  9892 &  9880 & 9893(1)\\

  $\chi_{b1}(2P)$  &  10262  & 10246  &10255(1)\\

   $\chi_{b1}(3P)$  & 10540 &  10541 & 10512(2)~\cite{1}\\
                                 &              &               & 10580(20)~ \cite{41}\\
   $\chi_{b1}(4P)$  &  10772  & 10793  &  abs.\\

   $\Upsilon(1D)$&  10161  &  10141  &  10164(1)\\

   $\Upsilon(2D)$  & 10460 & 10440 &  abs. \\

   $\Upsilon(3D)$  & 10704  &10701  & abs. \\

   $\Upsilon(4D)$  &  10915&  10933  &  abs.\\

   $\Upsilon(5D)$  &  11105  &11145   &abs. \\
 \hline
 \end{tabular}
 \end{table}

Here we pay attention to the fact that this ERT predicts  the correct value of the mass of the
$\Upsilon(1\,^3D_1)$ state, while the solution of the SSE  is $\sim 20$\,MeV smaller. Also the  ERT gives the mass of
$\chi_{b1}(3P)$ between the values observed in the experiments of LHCb \cite{6} and Belle \cite{41}.

\section{The Regge trajectories in the $(E^2,n)$- and $(E^2,l)$-planes in charmonium}

In charmonium there are only three multiplets ($1S$, $2S$, and $1P$) below the $D\bar D$ threshold and
these experimental data do not allow to extract exact values of the $c$-quark mass as well as all parameters of the
generalized ERT, Eq.~(\ref{eq.9}); nevertheless two mass differences, known from experiment,
\be
 M(\psi(2S)) - M(J/\psi) = 0.589(1)\,{\rm GeV}, ~M(\chi_{c1}(1P)) - M(J/\psi) = 0.414(1)\,{\rm GeV},
\label{eq.10}
\ee
put restrictions on the parameters of the ERT. Our analysis shows that  the main uncertainty comes from
a variation of the $c$-quark mass, entering the relation Eq.~(\ref{eq.3}).
Varying $m_c$ in the range $(1.2-1.4)$~GeV, the best agreement in the description of the
charmonium spectrum is  reached for  $m_c=(1.22-1.28)$\,GeV. Note that this value of  $m_c$ coincides with the current $c$-quark mass,
$\bar{m}_c(\bar{m}_c)=1.26(6)$~GeV \cite{35,36,37}. Here we take $m_c=1.24$~GeV. Then by definition,
\be
 a(\psi(nS)) = (M(J/\psi) - 2.48\,{\rm GeV})^2 = 0.3807\,{\rm GeV}^2,
\label{eq.11}
\ee
while the slope of the radial ERT is extracted from the mass difference, $M(\psi(2S)) - M(J/\psi)=0.589$~GeV,
\be
\sqrt{a + b_n} - \sqrt{a} = 0.589\,{\rm GeV},
\label{eq.12}
\ee
which gives
\be
 b_n(l=0) = 1.0738\,{\rm GeV}^2,\quad (m_c=1.24\,{\rm GeV}).
 \label{eq.13}
\ee
This value is smaller than the radial slope of the $\rho(nS)$-trajectory, $\beta_n(n\bar n,l=0)=1.45(5)$~GeV$^2$, but close to the value of the radial slope
$\beta_n(n\bar n,l\not= 0)=1.14(3)$\,GeV$^2$ for the $a_1(nP)$ and $\rho(n\,^3D_1)$ trajectories in light mesons.

Correspondingly, from the mass difference $M(\chi_{c1}(1P)) - M(\psi(1S))=\sqrt{a + b_l} - \sqrt{a}=0.414$\,GeV the slope of the orbital ERT,
\be
 b_l(m_c=1.24\,{\rm GeV}) =0.6863\,{\rm GeV}^2,
\label{eq.14}
\ee
is extracted, which  appears to be significantly smaller than the slope $b_l(n\bar n)=1.1(1)$~GeV$^2$ in light mesons.

Then all charmonium states (with the exception of the $\chi_{c0}(nP)$, see below) are described by the renormalized ERT,
\be
  E^2(c\bar c) ({\rm in~GeV}^2) = 0.3807  + 1.0738 \, n + 0.6863\, l, ~ m_c=1.240\,{\rm GeV},
  \label{eq.15}
\ee

The ERT of  $\chi_{c0}(nP)$ needs to be considered separately, because of  its large fine-structure splitting, which gives a smaller
mass difference $M(\chi_{c0}(1P)) - 2 m_c=0.318$\,GeV than  the value  0.414 GeV for $\chi_{c1}(1P)$. Therefore  a smaller
slope $b_l(\chi_{c0})=0.4935$\,GeV$^2$ is extracted and for  $\chi_{c0}(nP)$ its generalized ERT  is
\be
 E^2(\chi_{c0}(nP))({\rm in~GeV}^2)  = 0.3807 + 1.070 \,n + 0.4935\, l.
 \label{eq.16}
\ee
In Table~\ref{tab.3} we give the masses  $M(c\bar c,nl)= 2 m_c + E(c\bar c)$  of the states  $\psi(nS)$,
$\chi_{c1}(nP)$,  $\chi_{c0}(nP)$, and $\psi(n\,^3D_1)$, calculated according to the ERT and also the
solutions of the  SSE (including the fine-structure corrections).
\begin{table}
\caption{The masses of $\psi(nS),~\chi_{c1}(nP)$,  and $\psi(n\,^3D_1)$ (in MeV)\label{tab.4}}
\begin{tabular}{|c|c|c|c|}
\hline
 State    & SSE Eq.~(\ref{eq.7}) & ERT Eqs.~(\ref{eq.15},\ref{eq.16}) & experiment \\
\hline
$J/\psi$       & 3100     & 3097    & 3097  \\
$\psi(2S)$     & 3685    & 3686    &  3686 \\
$\psi(3S)$    &4100     &  4070    &  4039(1) \\
$\psi(4S)$    & 4455   & 4378     & 4346(6) or  4421(4)    \\
$\psi(5S)$    & 4760  & 4642      &4643(9) \\
$\psi(6S)$    & 5043    &4878   & abs.  \\

$\chi_{c1}(1P)$ & 3500  & 3513   &  3.510.7(1)  \\
$\chi_{c1}(2P)$  & 3949 &  3943  &  3871.7(2)   \\
$\chi_{c1}(3P)$ &4319  &  4273  &  4274(8)\\
$\chi_{c1}(4P)$  & 4642  & 4549 &  abs.  \\
$\chi_{c1}(5P)$&  4933   & 4796  &  abs. \\
$\chi_{c1}(6P)$  & 5201  & 5017  &  abs.  \\
$\chi_{c0}1P)$    & 3435 &   3415  &  3414.8(3) \\
$\chi_{c0}(2P)$   & 3929    &  3876   &   3918(2)\footnote{ not yet identified  in PDG as
$\chi_{c0} (nP)$ resonance} \\
$\chi_{c0}(3P)$   & 4289     &  4218  &  abs.   \\
$\chi_{c0}(4P)$   & 4622     & 4504 & 4506(25)$^a$\\
$\chi_{c0}(5P)$   & 4920     & 4754  & $4704^{+24}_{-38}$ $^a$ \\
$\psi(1\,^3D_1)$   & 3802  & 3804   &3773.1(4)  \\
$\psi(2\,^3D_1)$   & 4.188   &  4161  &  4191(5) \\
$\psi(3\,^3D_1)$   & 4.521 & 4455   & abs.  \\
$\psi(4\,^3D_1)$   & 4.821  & 4710  &   abs. \\
$\psi(5\,^3D_1)$    & 5095 &  4939  &  abs. \\
\hline
\end{tabular}
\end{table}

From Table~\ref{tab.4} one can see that in our  calculations  the masses of high excitations with $J^{PC}=0^{++}$,
$M(\chi_{c0}(3P))=4504$\,MeV  and $M(\chi_{c0}(4P))=4754$\,MeV,  are close to those of the $X(4500)$
and  $X(4700)$ resonances with $J^{PC}=0^{++}$, observed by the LHCb Collaboration \cite{6}. This coincidence can be
considered as an indication that these resonances could have a large $c\bar c$ component. Notice that due a large fine-structure shift down of
$\chi_{c0}(1P)$ ($\approx - 110$~MeV), the corresponding ERT has a a smaller orbital slope,
$b_l(\chi_{c0})=0.4935$~GeV$^2$, than that of $b_l(\chi_{c1})=0.6863$~GeV$^2$.

In Table~\ref{tab.4} we give also the masses, defined as the solutions of the SSE plus spin-dependent corrections,
where in the static potential $V_0(r)$ the linear confining potential is taken at all distances, i.e.,  the flattening effect at large distances was neglected.
For that reason the higher $nl$ resonances with $n=3,4,5$, determined by the SSE, appear
to be by  $(100-200)$\,MeV larger than those defined by the ERT. This interesting fact shows that in the ERT the flattening effect
is taken into account.

It is also worth to underline that the ERTs present the physical picture in HQ in a clear way and one can see how the ERT parameters
change when decreasing  the quark mass and, moreover, in charmonium the radial slope is already almost equal to that in light
mesons\cite{42}, where the generalized ERT,
\be
 M^2(nl,n\bar n) ({\rm in~GeV}^2) = 0.60 + 1.13(5) (n + l),~(l\not= 0)~m_q=0,
 \label{eq.17}
\ee
was obtained. Notice that in charmonium the radial slope of the ERT, $b_n(c\bar c)=1.0738$~GeV$^2$ is about three times
smaller than the one in the conventional radial RT with a given $l$,
\be
 M^2(c\bar c,n) = M^2(n=0) + \mu^2 n,
\label{eq.18}
\ee
where the radial slope, $\mu^2(c\bar c)\sim (2.8 -3.5)$\,GeV$^2$, has large value  \cite{26,27,28,29}.

\section{Discussion and Conclusions}

We have studied the HQ spectra in two approaches: with the use of the relativistic Hamiltonian with the universal interaction and via the
generalized ERTs, defined by the excitation energy, $E(nl)=M(nl)-2m_Q$. We have shown that the orbital and the radial ERT have different slopes, both in charmonium and bottomonium, and in bottomonium (charmonium) the intercept of the radial and the orbital ERT is the same, being smaller in bottomonium.  This fact allows to introduce the generalized ERT, Eq.~(\ref{eq.9}), which determines the HQ masses of the large number of states with $l=0,1,2,3$. The parameters of the ERT, as well as
$m_Q$,  were extracted from the experimental mass differences and their values are collected in Table~\ref{tab.5} together with those of the $\rho(\,^3S_1)$ and
$\rho(\,^3D_1)$ trajectories \cite{42}.

From Table~\ref{tab.5} one can see how the intercept, the orbital and the radial slopes are increasing  with a decreasing quark mass.
\begin{table}
\caption{The parameters of the generalized ERT (in GeV$^2$) in bottomonium,  charmonium and light vector mesons\label{tab.5}}
\begin{tabular}{|c|c|c|c|c|}
\hline
meson     & quark mass (in GeV) & intercept $a$ & orbital slope $b_l$  & radial slope $b_n$    \\
\hline
bottomonium    &     4.5492          & 0.1307           &     0.50      &  0.7242 $(l=0)$  \\
bottomonium    &                     &                  &               &  0.7060 $(l\not= 0$)\\

$\psi(^3S_1),~\psi(^3D_1),~\chi_{c1}(nP)$&  1.24  &  0.3807 &   0.6863  &  1.0738  \\

$\chi_{c0}(nP)$  &        1.24           & 0.3807       &  0.4935     &  1.070  \\

$\rho(^3S_1)$   & 0                &  0.60         &    0           &  1.45 (5) \\

$\rho(^3D_1)$    &  0              & 0.60       &   1.13(1)       &  1.13 (1)  \\
\hline
\end{tabular}
\end{table}
In bottomonium the resonances $\chi_{b1}(4P)$ with the mass 10756\,MeV and $\Upsilon(3\,^3D_1)$ with the mass 10700\,MeV
are predicted, while in charmonium the masses of the resonances $\chi_{c0}((n+1)P)$ with $J^{PC}=0^{++}$ and $n=3,4$, equal to 4504\,MeV, 4754\,MeV
respectively, are obtained. These masses appear to be close to those of
the $X(4500)$ and $X(4700)$ resonances \cite{6} and this fact can be considered as an indication that $X(4500)$ and $X(4700)$
have a large $c\bar c$ component in their wave function.

\begin{acknowledgements}
A.M.Badalian is very grateful to  Yu.A.Simonov for discussions.
\end{acknowledgements}

\end{document}